\documentclass[acmsmall]{acmart}

\usepackage[ruled,vlined]{algorithm2e}

\AtBeginDocument{%
  \providecommand\BibTeX{{%
    \normalfont B\kern-0.5em{\scshape i\kern-0.25em b}\kern-0.8em\TeX}}}

\begin{document}

\title[Botnets Breaking Transformers]{Botnets Breaking Transformers: Localization of Power Botnet Attacks Against the Distribution Grid}

\author{Lynn Pepin}
\email{lynn.pepin@uconn.edu}
\affiliation{%
  \institution{University of Connecticut}
  \country{}
}

\author{Lizhi Wang}
\email{lizhi.wang@stonybrook.edu}
\affiliation{%
  \institution{Stony Brook University}
  \country{}
}

\author{Jiangwei Wang}
\email{jiangwei.wang@uconn.edu}
\affiliation{%
  \institution{University of Connecticut}
  \country{}
}

\author{Songyang Han}
\email{songyang.han@uconn.edu}
\affiliation{%
  \institution{University of Connecticut}
  \country{}
}
  
\author{Pranav Pishawikar}
\email{pranav.pishawikar@uconn.edu}
\affiliation{%
  \institution{University of Connecticut}}

\author{Amir Herzberg}
\email{amir.herzberg@uconn.edu}
\affiliation{%
  \institution{University of Connecticut}
  \country{}
}

\author{Peng Zhang}
\email{P.Zhang@stonybrook.edu}
\affiliation{%
  \institution{Stony Brook University}
  \country{}
}

\author{Fei Miao}
\email{fei.miao@uconn.edu}
\affiliation{%
  \institution{University of Connecticut}
  \country{}
}



\begin{abstract}
Traditional botnet attacks leverage large and distributed numbers of compromised internet-connected devices to target and overwhelm other devices with internet packets. But with increasing consumer adoption of high-wattage internet-facing ``smart devices", a new ``power botnet" attack emerges, where such devices are used to target and overwhelm power grid devices with unusual load demand. We introduce a specific variant of this attack, the \emph{power-botnet weardown-attack}, which does not intend to cause blackouts or short-term acute instability, but instead forces expensive mechanical components to activate more frequently, necessitating costly replacements or repairs. Specifically, we target the on-load tap-changer (OLTC) transformer, which involves a mechanical switch that responds to change in load demand. In our analysis and simulations, such power botnets can halve the lifespan of an OLTC, or in the most extreme cases, reduce it to $2.5\%$ of its original lifespan. Notably, these power botnets are composed of devices that \emph{are not connected to the internal SCADA systems} used to control power grids. This represents a new internet-based cyberattack that targets the power grid in a way that cannot be solved by hardening existing SCADA systems. To help the power system to mitigate these types of botnet attacks, we develop attack-localization strategies. To the best of our knowledge, there is no valid model-based approach for attack localization. So we formulate the problem as a supervised machine learning task to locate the source of power botnet attacks. Within a simulated environment, we generate the training and testing dataset to evaluate several machine learning algorithm based localization methods, including SVM, neural network and decision tree. We show that decision-tree based classification successfully identifies power botnet attacks and locates compromised devices with at least $94\%$ improvement of accuracy over a baseline ``most-frequent"  classifier.
\end{abstract}

\ccsdesc[500]{Security and privacy~Hardware attacks and countermeasures}
\ccsdesc[500]{Computing methodologies~Machine learning}
\ccsdesc[300]{Computing methodologies~Machine learning algorithms}
\ccsdesc[100]{Computing methodologies~Classification and regression trees}
\ccsdesc[100]{Computing methodologies~Supervised learning}
\ccsdesc[500]{Hardware~Smart grid}
\ccsdesc[500]{Hardware~Power networks}
\ccsdesc[300]{Hardware~Switching devices power issues}

\keywords{power system, distribution grid, machine learning, supervised learning, decision tree, IEEE 123, OLTC, power botnet, botnet}

\maketitle

\section{Introduction} \label{sec:intro}

The rapidly-advancing field of artificial intelligence (AI) has been increasingly applied to address the security issues in cyber-physical systems (CPS).  Numbers of machine learning methods, including support vector machine (SVM), multinomial logistic regression (LR), random forest (RF), naive bayes (NB) and multi layered perceptron (MLP) neural networks have been applied in malware detection~\cite{li2018significant}, in intrusion detection~\cite{junejo2016behaviour}, and in attack and faults differentiation~\cite{tertytchny2019differentiating}. Deep learning methods, such as convolutional neural networks (CNN), long short-term memory (LSTM) and autoencoders (AEs), have also played important roles in classification and detection of malware variants~\cite{kolosnjaji2016deep,wang2019effective}, anomaly detection~\cite{liu2017survey}, and malicious node detection\cite{article} in CPS.

One of the many attacks employed against internet-connected systems is the \emph{botnet attack}. A \textbf{botnet} is a network of compromised internet-facing devices (called \textbf{bots}) which are controlled and coordinated through the internet by an attacker~\cite{li2009botnet}. These botnets are usually composed of compromised devices such as personal computers, security cameras, routers, or printers.  Botnets are usually used for malicious purposes, and they operate by leveraging their large and distributed numbers of internet-connected devices. Their strength can typically be characterized by their total upload bandwidth (such as for launching distributed denial of service (DDoS) attacks) or their total computational power (such as for mining cryptocurrencies.)

The increasing consumer adoption of poorly-secured IoT devices also means a growth in potential bots, making IoT devices an attractive target for malicious hackers~\cite{oconnor2019blinded}. A concerning trend is that consumers are increasingly purchasing internet-connected versions of certain high-wattage devices such as refrigerators or water heaters. This is concerning because, in addition to networking and computational resources, the attacker gains a new capability of the botnet: The ability to manipulate a portion of power grid load demand.

Recent literature introduces this new type of botnet related to power systems and the above mentioned high-wattage IoT devices. 
We call such a botnet a \textbf{power botnet}. If coordinated by an attacker, the devices (which are used at the homes and offices at the edge of the distribution grid) can raise and lower their power consumption to cause instability in the power system. Alone, a single consumer device represents a tiny fraction of power consumption of a distribution network, and is unlikely to cause any instability.  But a large number of compromised devices distributed throughout the distribution grid acting in coordination can cause impactful instabilities. Such attack scenarios have been explored in literature for the transmission network of power systems. A plausibly-sized power botnet might provide an attacker enough control over the demand of a power grid to cause power outages, blackouts, and other immediate damage~\cite{soltan2018blackiot}. One result shows that between 2.5 million and 9.8 million infected desktop computers are needed to attack the synchronous grid of Continental Europe~\cite{dabrowski2017grid}. The attacks proposed in these works exploit the assumption that load demand is predictable. By deviating from the expected load, the power grid is quickly made unstable. 

Notably, these power botnet attacks do not take advantage of any security flaws in the power grid Supervisory Control and Data Acquisition (SCADA) system. All the damage to the power system in these attacks can be done without any compromise of the SCADA system, and the security of the IoT devices is outside the scope of control of the grid operators. This is important, because these attacks represent a cybersecurity threat that grid operators can not mitigate by traditional cybersecurity techniques alone. 

Several recent works have proposed different strategies to mitigate and protect power grids against these power botnet attacks. Two algorithms, SAFE and IMMUNE, are devised to find the robust operating point for generators to avoid line overloads under power botnet attacks~\cite{soltan2018protecting}. A scalable mitigation approach is proposed to detect and mitigate false data injection (FDI) attacks and DoS attacks initiated by compromised IoT devices~\cite{yilmaz2019timely}.  Methods to mitigate and protect against power botnet attacks designed to cause acute instability (blackouts, etc) have been studied~\cite{soltan2018protecting}.

\textcolor{black}{Further work suggests that the risk that power botnets pose to the transmission network might already be mitigated by existing protection mechanisms~\cite{huang2019not}, suggesting that immediate blackouts might actually be very difficult to achieve. Other instabilities, such as bringing power frequencies outside safety threshholds, are considered more easily attainable.}

\textcolor{black}{However, all these works explore attacks using power botnets to cause acute instabilities. Frequency instability or power outages will have immediate and damaging impacts, but these inherently self-limit the attack, and the immediate, obvious impacts will bring investigation and attention to the attacker. Rather than focus on causing acute instability, we instead consider ``weardown attacks". That is to say, we consider  long-term power-botnet attacks that cause expensive components of the power infrastructure to degrade faster than is normal. Our work focuses on wearing down the the ``on load tap changer" (OLTC) transformer, an expensive component of the power system that uses a mechanical switch to respond to changes in demand.}


\textcolor{black}{To summarize, motivations of our work are illustrated as following:
\begin{itemize}
  \item The increasing adoption of high-wattage IoT devices mean an increasing amount of consumer power demand can be considered an internet-controlled load. This has the potential to fall under attacker control, as a power-botnet. These attacks become increasingly more plausible.
  \item The impact of power-botnet attacks have been explored in recent literature, but the majority of work has focused on attacks with immediately obvious impacts, such as blackouts or frequency instability. Instead, we consider attacks that degrade the longer-term health of the system.
  \item OLTC transformers are being increasingly adopted in distribution grids, particularly in the context of residential-scale PV systems \cite{reese2012enhanced, wang2014integrating}. The OLTC transformer is one of the most important and expensive devices in power systems.
  \item Even if the IT security of the SCADA control systems were perfect, power botnets at the grid edge pose a unique challenge. The security of these high-wattage IoT devices is outside the grid operators control, and the threat of the power-botnet is not a typical network or computer security threat.
  \item The main cost to the attacker is in amassing vulnerabilities against devices. Once done, these attacks can be deployed easily and the economic impact is multiplied for every OLTC. If not detected, this would cause widespread early retirement of OLTC transformers.
\end{itemize}}

In this work, we focus on locating power botnets during attacks in the IEEE 123 bus test feeder. Attack mitigation and analysis on other power systems will be our future work. To the best of our knowledge, no model-based approach exists that will allow us to perform localization using closed-form equations. So we approach localization using learning methods. 
Whereas traditional botnet detection and mitigation techniques employ some level of network traffic analysis~\cite{tzagkarakis2019botnet, doshi2018machine}, in this work we focus only on measurable power-grid information such as voltage, current, and phase angle to identify and locate devices in the power botnet. This is in line with existing CPS intrusion detection methods~\cite{li2018significant, junejo2016behaviour}. \textcolor{black}{In our work, we don't apply anomaly detection or other unsupervised learning methods, but we instead perform supervised learning by training a classification model to map the time-series power-grid information to attacker locations. Collecting labelled training data and repeating this work is feasible in real-world scenarios since grid operators can use a copy of their digital twin~\cite{tao2018digital} model to simulate attacks against.
The existing works usually deal with the immediate and obvious impacts like instability or blackouts, therefore, it is easy to detect those anomalies since there will be abrupt changes in the measured voltage, current, etc., from the feeder \cite{li2018fault, duan2016fault}. However, in our work, the attacker only controls the load to cause more tap changes. Instead of abrupt and remarkable changes in the power measurements, there will just be small changes caused by the changing loads. Therefore, It's hard to detect the attack immediately just from the short-term measurements since grid operator cannot know whether there are attackers or consumers changing the loads. }

In our previous work, we have developed attacks which cause long-term hardware damage in the power grid~\cite{wang2018powerbotnet}. These attacks target the on-load tap-changer (OLTC) component of transformers. By fluctuating the controlled load demand, the physical switch inside the OLTC moves more frequently than is usual, reducing the lifespan of the device. We introduced an OpenDSS simulation~\cite{montenegro2012real} utilizing the CREST~\cite{mckenna2016high} household power consumption data to simulate these power botnet attacks against OLTC mechanisms inside the IEEE 123 test system~\cite{subcommittee2004ieee, 8063903}. We expand and continue with this simulator for the evaluation section of this work.

In particular, the contributions of this work are as follows.

\begin{itemize}
\item We are the first to describe and analyze \emph{weardown attacks} utilizing power botnets, where other works focus on causing acute instability.
\item We analyze three power botnet attacks algorithms targeting the OLTC system, and show how these attacks impact the longevity of the OLTC mechanism as a function of the attacker strength. 
\item We design a supervised decision tree based learning method to instantaneously locate the nodes with compromised power botnet devices. With no tuning, the decision tree outperforms both best-effort neural network and linear SVM classifiers, with normalized-accuracies between 94\% and 99\% across the thirteen nodes.
\end{itemize}

The rest of this paper is organized as follows: In Section~\ref{sec:attack_problem} we describe the power system model and power botnet attack problem formulation for an attacker desiring to discretely decrease the lifespan of an OLTC. In Section~\ref{sec:attack_method}, we propose three different strategies for a successful attack and analyze their efficacy at decreasing the lifespan of an OLTC. In Section~\ref{sec:learning_problem}, we define the goal of locating nodes containing devices in the power botnet and the machine learning method based attack detection and localization problem formulation. In Section~\ref{sec:learning_metroid}, we design a decision tree learning method and analyze its performance under three different metrics: normalized accuracy, true positive rate (sensitivity), and true negative rate (specificity). In Section~\ref{sec:conclusion} we summarize and conclude this paper.
\section{System Model and Attack Problem Formulation} \label{sec:attack_problem}

\begin{figure}
  \centering
  \includegraphics[width=.5\textwidth]{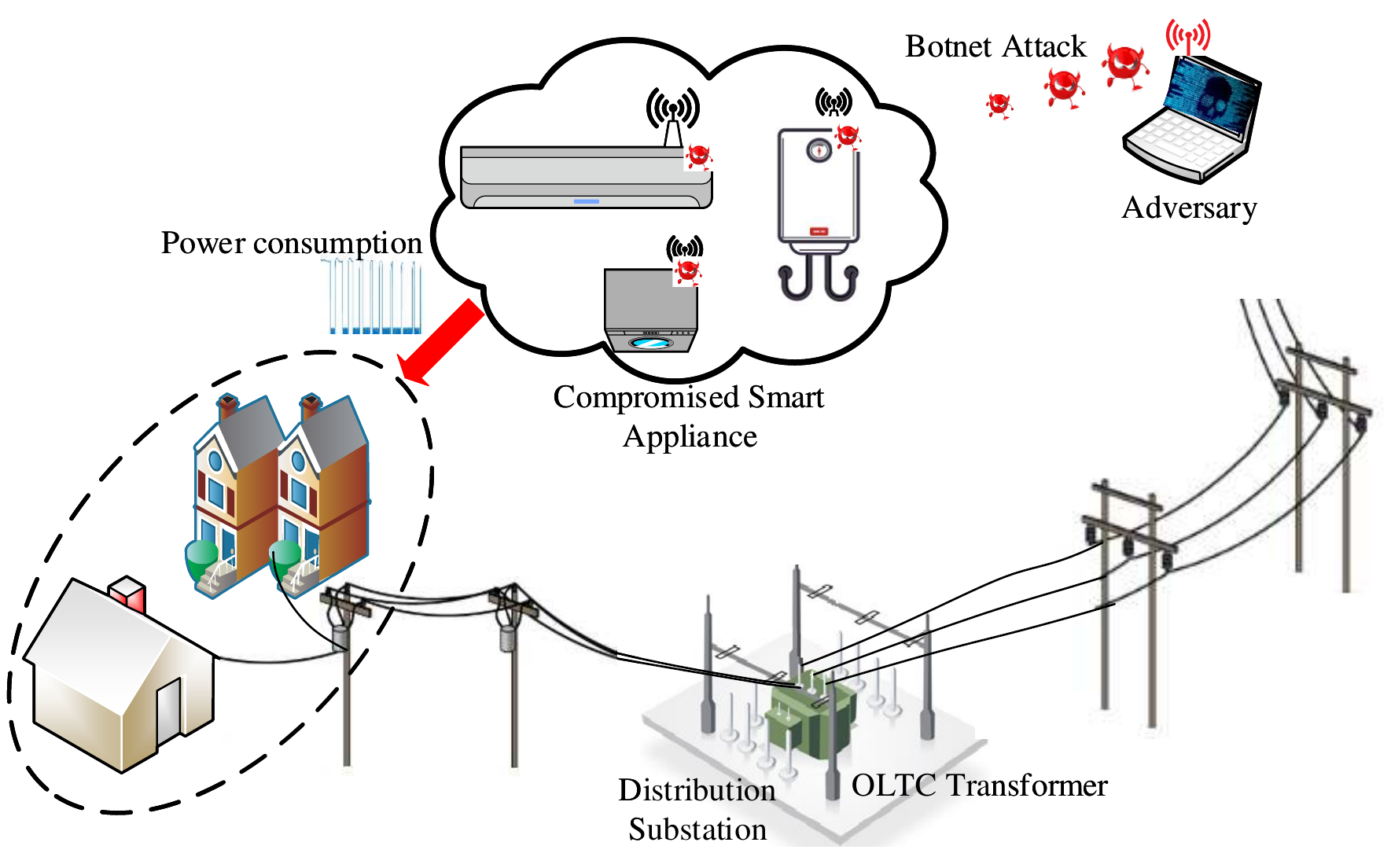}
  \caption{The Power Botnet Attack} 
  \label{fig:pbnet_fig1}
\end{figure}

\begin{figure}
  \centering
  \includegraphics[width=.8\textwidth]{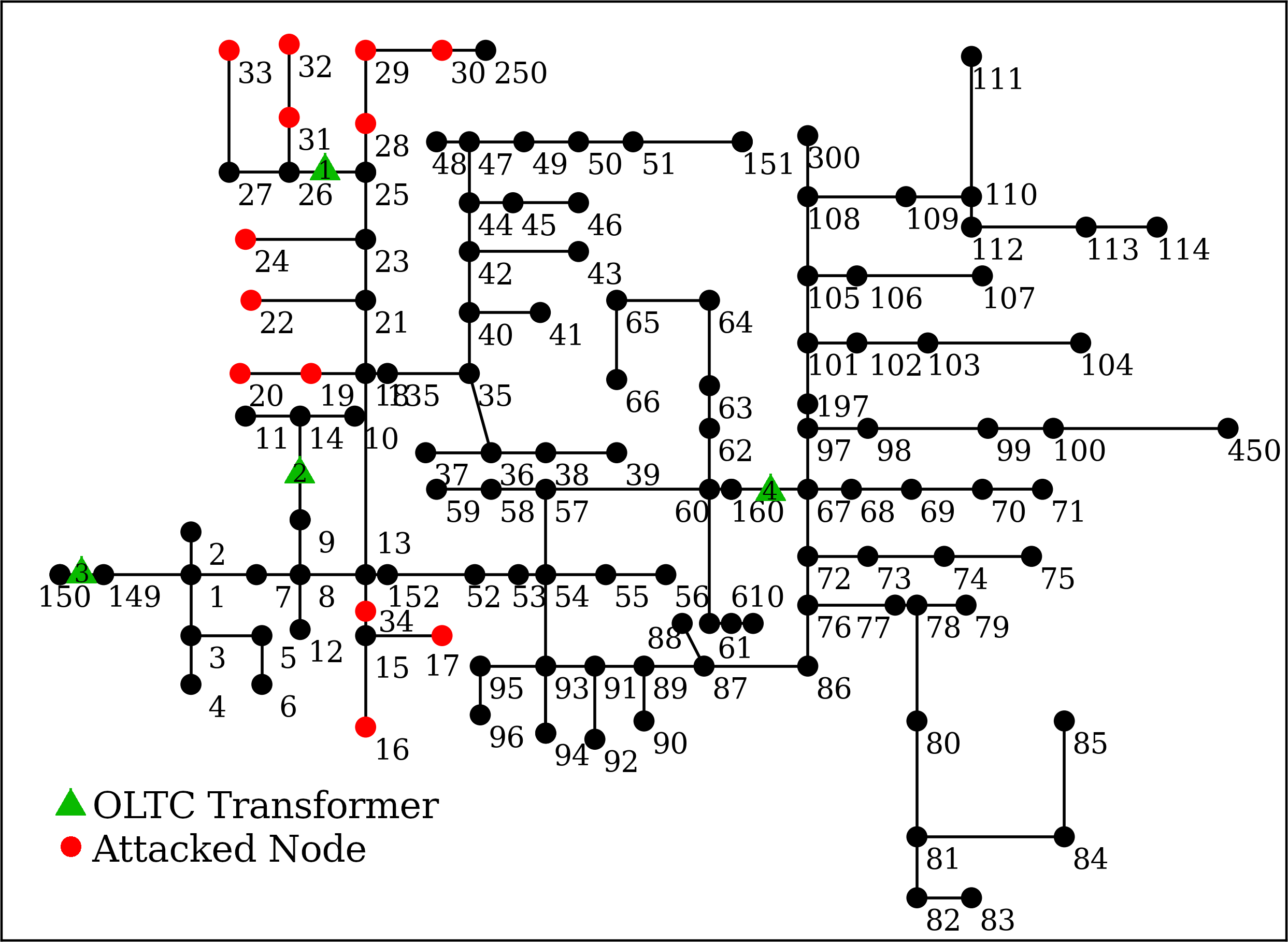}
  \caption{IEEE 123 bus system involving attacked areas}
  \Description[The IEEE 123 bus system, with targeted nodes highlighted in red.]{The IEEE 123 bus system is represented as a grid of connected nodes, labeled with numbers 1 through 114, plus other nodes labeled 135, 149, 150, 151, 160, 250, 300, 350, 450, 451, 610. There are also four triangular nodes throughout the graph, labeled with numbers 1, 2, 3, and 4, representing the four OLTC. Thirteen nodes are labeled red, as noted in the paper.}
  \label{fig:IEEE123}
\end{figure} 

In this section, we describe the simulation setup as well as the attacker goals and capabilities. The structure of this attack is referenced in Figure~\ref{fig:pbnet_fig1}.

We utilize the OpenDSS simulator~\cite{montenegro2012real} for the power simulation and we utilize the topology of the IEEE 123 node test feeder~\cite{subcommittee2004ieee, 8063903}, as depicted in Figure~\ref{fig:IEEE123}. We choose this test feeder as a representative example scenario. Each node in the feeder represents four households, each with one waterheater. Daily household power consumption is provided for OpenDSS by utilizing the CREST tool~\cite{mckenna2016high}. A given simulation scenario is one day long, producing data for $T = 1440$ discrete timesteps, with each timestep representing one-minute. 

Within the IEEE 123 test system, there are four labelled OLTC mechanisms. For simplicity, we focus only on OLTC 1, which exists between nodes 26 and 25. We identify thirteen nodes which may contain compromised devices. These thirteen nodes are labelled 16, 17, 19, 20, 22, 24, 28, 29, 30, 31, 32, 33, and 34, as shown by the red nodes in Figure~\ref{fig:IEEE123}. Each node has an associated $90 kW$ of manipulable load (with each node representing four households), except for nodes 17, 31, and 32, which have $45 kW$ of manipulable load (representing two households each).

For a given simulation scenario, an attacker's capabilities are described by the ``attacker load ratio" value $ALR$, and which combination of the thirteen nodes it controls. We define the ``attack load ratio" to approximately measure what percentage of the load demand the attacker can influence, compared to usual load. For a given node, the attacker can draw  up to $90 kW * ALR * \frac{13}{\text{total controlled nodes}}$ of load, substituting $45 kW$ with nodes 17, 31, and 32.

\subsection{Attacker Capabilities and Objective}

To repeat, for a given simulation scenario an attacker's capabilities may be described by the value of $ALR$ and by which combination of the thirteen nodes the attacker controls. An attacker is also described by the attack strategy it chooses. An attacker is permitted to make use of any piece of information from the simulated power system, such as voltage, current, or power consumption values at given nodes. The model also assumes no internet delay or instability, meaning the attacker can synchronously and instantly read any piece of information or activate/deactivate given devices. 

The ultimate goal of the attacker is to lower the lifespan of these OLTC mechanisms, or equivalently, to maximize the number of tap changes. The attacker is limited by the number of timesteps $T=1440$ in a given simulation. This means there are a maximum of $1440$ possible tap changes in a simulated day, up from the nominal value of $36$ taps. This is a 40 times increase in tap changes, or equivalently, a reduction in lifespan to $2.5\%$ of its original lifespan. Precisely speaking, at each timestep, an attacker is able to observe the entire state of the power grid, and then can activate or deactivate any of the thirteen nodes under its control. This attacker's choice may be represented by a vector of dimensionality $13$ taking values in the range $0$ to $1$.

\textcolor{black}{Because these devices are controlled by attackers, we make the implicit assumption that any sensors or network devices outside of the grid operators control are also compromised by attackers. That is to say, we can not simply perform load analysis \emph{on the IoT device itself} as we assume the attackers are capable of disabling or modifying that as well. This motivates our localization strategy to be limited to only read data from the OLTC transformer, system busses, and other `internal' measurement points.}

\subsection{Sample IoT Device: Water Heater Model}


\label{tab:water-heater}
\begin{table}[hb]
\caption{Water Heater Table of Notation}
\begin{center}
\begin{tabular}{r p{10cm} }
\toprule 
    $Q_{wh}(t)$     & Thermal energy in water heater over time \\
    $M_{wh}$          & Mass of waterheater, constant, $M_{wh} = 172.2kg$ \\
    $C$             & Specific heat capacity of water, constant, $C = 4181.3 \frac{J}{kgK}$\\
    $h$             & Heat-transfer coefficient, constant, $h = 0.9142 \frac{J}{m^2Ks}$ \\
    $A_{wh}$        & Surface area of waterheater, constant, $A_{wh} = 149.7m$ \\
    $T_a$           & Ambient temperature around waterheater, constant, $T_a = 293.15K$ \\
    $T_i$           & Initial temperature of waterheater, constant, $T_i = 293.15K$ \\
    $P_{wh}(t)$     & Power consumption of waterheater at time $T$, measured in watts ($W$, or $J/S$) \\
    $ALR$           & Value in the range $[0, 1]$.\\
\bottomrule
\end{tabular}
\end{center}
\end{table}
\textcolor{black}{To define the capability of the attacks, we propose a simple model for a home water heater, with limited heat capacity of that can limit the power consumption capacity of the attacker by manipulating one device. These water heaters represent a manipulable load in the system, for example, in the scenario that a widely-used waterheater had a vulnerability discovered by our attacker. The waterheater model is one example of feasible design, and a model of vulnerable IoT device that can help to explain the design of an attack strategy and capability, and the corresponding attack strategies can be simulated in OpenDSS. When the attackers can manipulate other devices, similarly, with knowledge about the load and power consumption the attacker can control, we can simulate the process and analyze the impact.}

For instance, we associate a waterheater to each household, which represents a load that may be compromised and utilized by an attacker. When active, a waterheater adds thermal energy to a body of water, limited by a certain temperature threshhold. We model a waterheater with the following differential equation, based off Newton's Law of Cooling,

\begin{equation}\label{eq:water-heater-de}
    \frac{d}{dt}Q_{wh}(t) = P_{wh}(t) - hA_{wh}\left(\frac{Q_{wh}(t)}{M_{wh}C}-T_a\right)
\end{equation}

where the current temperature of the waterheater is given in the term $\frac{Q_{wh}(t)}{M_{wh}C}$, and with each item described in Table~\ref{tab:water-heater}. This differential equation yields the following analytic form for $Q_{wh}(t)$:

\begin{equation}\label{water-heater}
    Q_{wh}(t) = M_{wh}C\left(\frac{P}{hA_{wh}} + T_a + \left(T_i - T_a - \frac{P_{wh}(t)}{hA_{wh}}\right)e^{-\frac{hA_{wh}}{M_{wh}C}t} \right)
\end{equation}


The constants used in this simplified waterheater model are derived in reference to~\cite{maguire2012parametric}. The waterheater represents the attacker's only influence over the power system. The strength of an attacker is described by the value $ALR$, where $ALR = 1.0$ describes influence over 100\% of the waterheaters in a given grid. The term $P_{wh}(t)$ takes on values between $0$ and $(4494 W)*ALR$. If the temperature of the waterheater $T(t) = Q_{wh}(t)/(M_{wh}C)$ exceeds $336 K$ (roughly $145F$ or $63C$), the waterheater can not be activated and $P_{wh}(t)$ takes on value $0 W$.

The attacker takes control of a waterheater through an internet-connected control unit, and is able to set the temperature at each 1-minute timestep. In this manner, the attacker is able to control the load the waterheater exerts on the power grid.


\subsection{Modeling and Control of OLTC}

The OLTC mechanism is a physical switch which moves in response to unpredictable shifts in power demand. These are more expensive than traditional transformer mechanisms but are useful to prevent temporary cuts in power. 

\begin{figure}
  \centering
  \includegraphics[width=.4\textwidth]{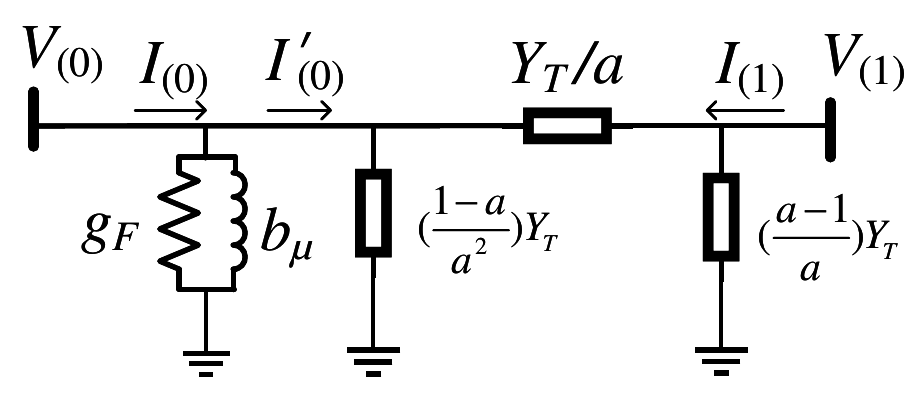}
  \caption{$\pi$-Equivalent circuit of OLTC}
  \Description[TODO]{TODO}
  \label{fig:eq circuit}
\end{figure}

\begin{equation}\label{e.OLTC1}
\left[
    \begin{array}{c}
    V_{(1,t)}\\I_{(1,t)}
    \end{array}
\right]
=\left[
        \begin{array}{cc}
        \frac{1}{a} &-\frac{a}{Y_T}\\
        0   &-a
    \end{array}
\right]
\left[
    \begin{array}{c}
    V_{(0,t)}\\I_{(0,t)}
    \end{array}
\right],
\end{equation}

Figure ~\ref{fig:eq circuit} shows the equivalent circuit model of OLTC, where $V_{(0)}, I_{(0)}, V_{(1)}, I_{(1)}$ represent primary, secondary voltage and current respectively.  The secondary voltage and current of OLTC are calculated as in~\cite{7005526} where $Y_T$ is the transformer series admittance, and $a$ is the turns ratio. To analyze the power flow of power distribution grid with OLTC, equation~\ref{e.OLTC1} can be rewritten as

\begin{equation}\label{e.OLTC2}
\left[
    \begin{array}{c}
    I_{(0,t)}\\I_{(1,t)}
    \end{array}
\right]
= \overbrace{\left[
    \begin{array}{cc}
    g_{Fe}+jb_\mu+\frac{Y_T}{a^2} &-\frac{Y_T}{a}\\
    -\frac{Y_T}{a}   &Y_T
    \end{array}
\right]}^\text{$Y_{OLTC}$}
\left[
    \begin{array}{c}
    V_{(0,t)}\\V_{(1,t)}
    \end{array}
\right],
\end{equation}

where $Y_{OLTC}$ is the OLTC admittance matrix, which represents the OLTC admittance in the power flow equations. The ratios can be given as \begin{equation}\label{e.OLTC3}
    a=a_0+n(t)\Delta a,
\end{equation}
 where $a_0$ is the nominal turns ratio which usually equals 1.0 p.u., $\Delta a$ is the tap change of $a_0$, and $n(t)$ is the tap position of OLTC at time $t$, which equals the sum of previous tap position and integar tap changes, given as \begin{equation}\label{e.OLTC4}
    n(t)=n(t-1)+\Delta n(t).
\end{equation}

\begin{figure}
  \centering
  \includegraphics[width=.5\textwidth]{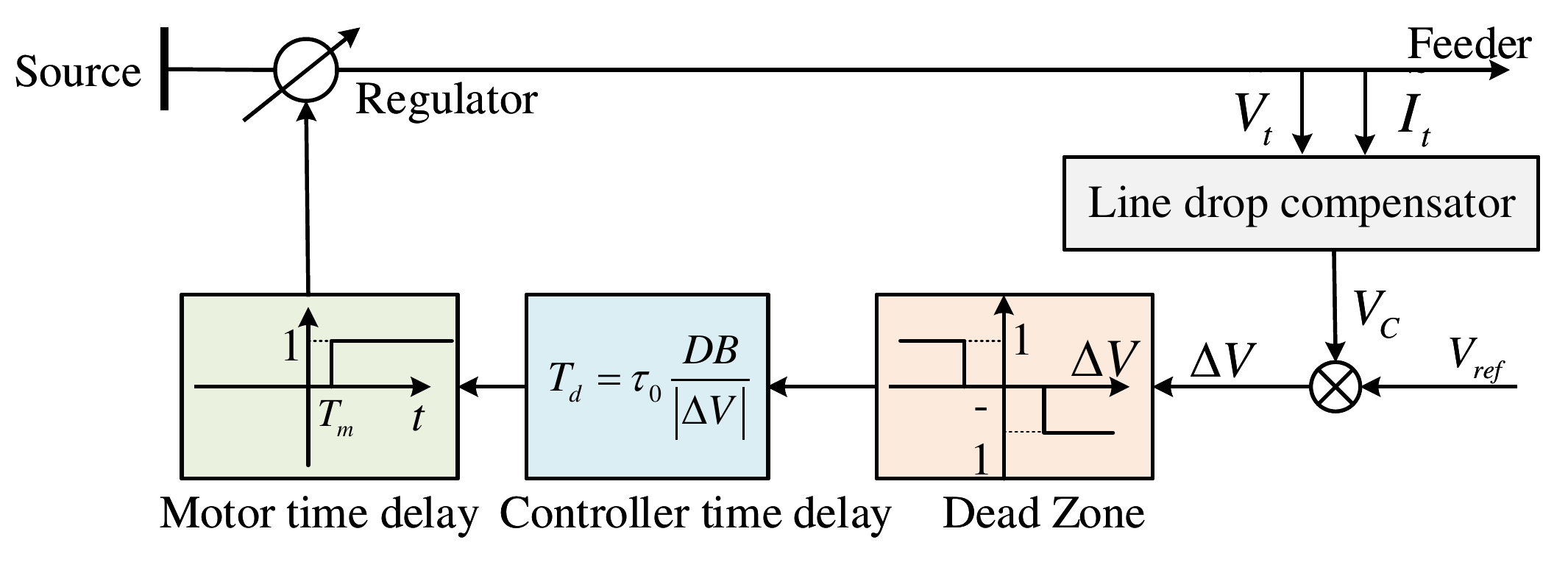}
  \caption{OLTC control block diagram}
  \Description[A block diagram showing mechanisms by which an OLTC decides to change tap.]{An incoming power source is passed through an OLTC. The modified power is both fed to the feeder and to the OLTC switching mechanism. The switching mechanism in the block diagram is described as follows. The value V sub c is calculated from the power source through the line drop compensator. This is compared to a reference voltage V sub ref to calculate Delta V. This Delta V value is passed through a Dead Zone function, the value of which is passed forward to a controller time delay, which is itself passed to a Motor time delay.}
  \label{fig:OLTCC}
\end{figure}

A typical tap position of an OLTC can be varied from $-{N_M}$ to $+{N_M}$ (inclusive) which constitutes $2N_M+1$ possible positions for each phase. The control diagram is shown in Figure~\ref{fig:OLTCC}, where $v_{ref}$ is the reference of regulated voltage, $\tilde{V_t}$ and $\tilde{I_t}$ are the voltage and current of node in the phasor mode and $V_C$ is the compensated voltage using the line droop compensator. Typically, the process of tap changing involves two time delays: a controller time delay $T_d$ which is intentionally introduced to avoid tap changing during fast voltage transients; and a mechanical time delay $T_m$ due to the motor drive mechanism of the OLTC. The mechanical time delay $T_m$ has a constant value, which usually varies from 3 to 10 seconds. The controller time delay $T_d$ is commonly considered as a variable time delay depending on the voltage error $\Delta V$, the constant values controller dead band \textbf{\textit{DB}} and $\tau_0$~\cite{1645174},

\begin{equation}\label{e.OLTC5}
    T_d=\tau_0\frac{DB}{\Delta V}.
\end{equation}

The control diagram of OLTC is a typical feedback control. The voltage of secondary side of OLTC will be feed back to the controller, the difference between which and set point will be used to control tap position of OLTC. To be specific, the abrupt increase of the load at the secondary side of OLTC will lead the feedback voltage of OLTC to decrease. Once the difference of the voltage is larger than the threshold, the tap will increase to maintain the feedback voltage. By controlling the loads at the secondary set, the attacker can easily cause the tap changes of OLTC. Details of the attack methods will be illustrated in the following section.

\section{Attack Methods and Impacts Analysis} \label{sec:attack_method}

In this section, we define three attack strategies considered in this work, and show their efficacy in reducing the lifespan of the targeted OLTC. In particular, we show three methods of attack: a randomized strategy, an alternative ``flipping" strategy, and a heuristic-based strategy. 

\textcolor{black}{To surmise, we consider an attack against the IEEE 123-bus test feeder, specifically a neighborhood of 13 nodes around a specific test feeder. Some subset of these thirteen nodes contain compromised devices under attacker control, and the attacker goal is to use these nodes to maximize the number of tap changes caused to the OLTC. The choice of the neighborhood of 13 nodes is just one example.}
At a given timestep, the attacker controls the devices under its control by sending a control vector of dimensionality $13$ with values in the range $0$ to $1$, where each value corresponds to one of the thirteen nodes. A value of $0$ indicates deactivating the devices in that node entirely, and a value of $1$ indicates activating the devices in that node to their maximum power consumption. Let $y_a \in \{0, 1\}^{13}$ be a 13-dimensional binary vector representing the thirteen attackable nodes, where the $i^{th}$ value of $y_a$ is $1$ if the attacker has control of the $i^{th}$ node and is $0$ otherwise. So, an attacker choosing vector $y_a$ is activating each node under its control, maximizing power consumption. 

\subsection{Description of Attack Strategies}
\paragraph{Random attack strategy} The random attack strategy is defined in Algorithm~\ref{alg:rand}. At each timestep, the attacker selects the nodes to be attacked as either $y_a$ or $0^{13}$ with a probability $p = 0.50$. The attack to all nodes is done synchronously, so either all nodes are turned on, or all nodes are turned off. The distribution grid relies on the assumption of a roughly predictable power demand. So in this method, the attacker performs as unpredictably as possible, flipping the devices on or off with probability $1/2$. To implement this attack, the attacker does not need to have additional knowledge about the dynamic of the system besides the basic assumptions of capabilities described in Section 2. Hence, this attack strategy is relatively naive. 

\begin{algorithm}[h]
\SetAlgoLined
\textbf{Input:} Probability $p$ such that $0 \leq p \leq 1$, $y_a$

Sample $q$ uniformly in the range $[0, 1]$.

\uIf{q < p}{
    Choose $y_a$, turning each node on.
}
\Else{
    Choose $0^{13}$, turning each node off.
}
\caption{Random-activation attack}
\label{alg:rand}
\end{algorithm}

\paragraph{Flipping attack strategy} The flipping attack strategy is defined in Algorithm~\ref{alg:flip}. This attack alternates between $y_a$ and $0^{13}$ each turn, synchronously turning each node on or each node off. This method is akin to ``flickering a lightswitch", with the intention being to cause rapid and large swings in powergrid demand.


\begin{algorithm}[h]
\SetAlgoLined
\textbf{Input: } Memory boolean $b$, $y_a$

\textbf{Output: } Updated value for $b$

\eIf{$b$}
{Choose $y_a$, turning each node on.}
{Choose $0^{13}$, turning each node off.}
$b = $ not $b$
\caption{Flipping attack}
\label{alg:flip}
\end{algorithm}

\paragraph{\textcolor{black}{Heuristic attack strategy}}

The heuristic attack strategy referenced in Algorithm~\ref{alg:aggr} is the most complicated of the three. The goal of this attack is to estimate if activating the loads will cause a tap change in the given timestep. If so, the attacker chooses $y_a$, activating every node synchronously, and then chooses $0^{13}$ the following turn to cause another tap change from the falling load.

\begin{algorithm}[h]
\SetAlgoLined
\textbf{Input: } $V_t$, $R_C$, $X_C$, $\tilde{I}_t$, $b$, $\epsilon$, $y_a$

\textbf{Output: } Updated value for $b$

$V_C(t)=|\tilde{V}_t+(R_C+jX_C)\tilde{I}_t|$

\uIf{$V_C(t) \leq \varepsilon $ and $b$}{
    Choose $y_a$, turning each node on.
    
    $b = $ False
}
\Else{
Choose $0^{13}$, turning each node off.

$b = $ True
}
\caption{Heuristic attack}
\label{alg:aggr}
\end{algorithm}

The heuristic strategy takes advantage of the mathematical mechanics of OLTC and known information about the power grid state. It was intended to be a `smarter' attack. The regulated voltage of OLTC fluctuates around the boundary $V_{ref}-\textbf{DB}/2$ (where $DB$ is a deadband value used for hysteresis). At each time step, the attacker calculates the feedback voltage~\cite{kundur1994power}, $V_C(t)$, of the OLTC as $V_C(t)=|\tilde{V}_t+(R_C+jX_C)\tilde{I}_t|$, and if it is less than a fixed parameter $\epsilon$, the attacker activates each device. For this strategy, a value of $\epsilon = 2397 V$ is chosen. The idea is to heuristically estimate when an increase in load would cause a tap change, and when such a scenario occurs, the attacker will activate the loads to force a tap change. This is called a ``rising" tap change. When the loads are activated in one time step, the attacker then deactivates loads in the next time step. This has the potential to cause an additional ``falling" tap change. Deactivating the waterheater also helps to preserve the limited thermal capacity of the water heater, preventing them from overheating.

\subsection{The Impacts of Attack Strategies on Lifespan}

The figures in this section show how the lifespan of the OLTC changes as a function of the attacker strength $ALR$ for each of the three strategies. For these results, we assume that the attacker has access to the IoT devices (water heaters) connected to the nodes selected to be attacked.

\begin{figure}[h]
  \centering
  \includegraphics[width=.8\textwidth]{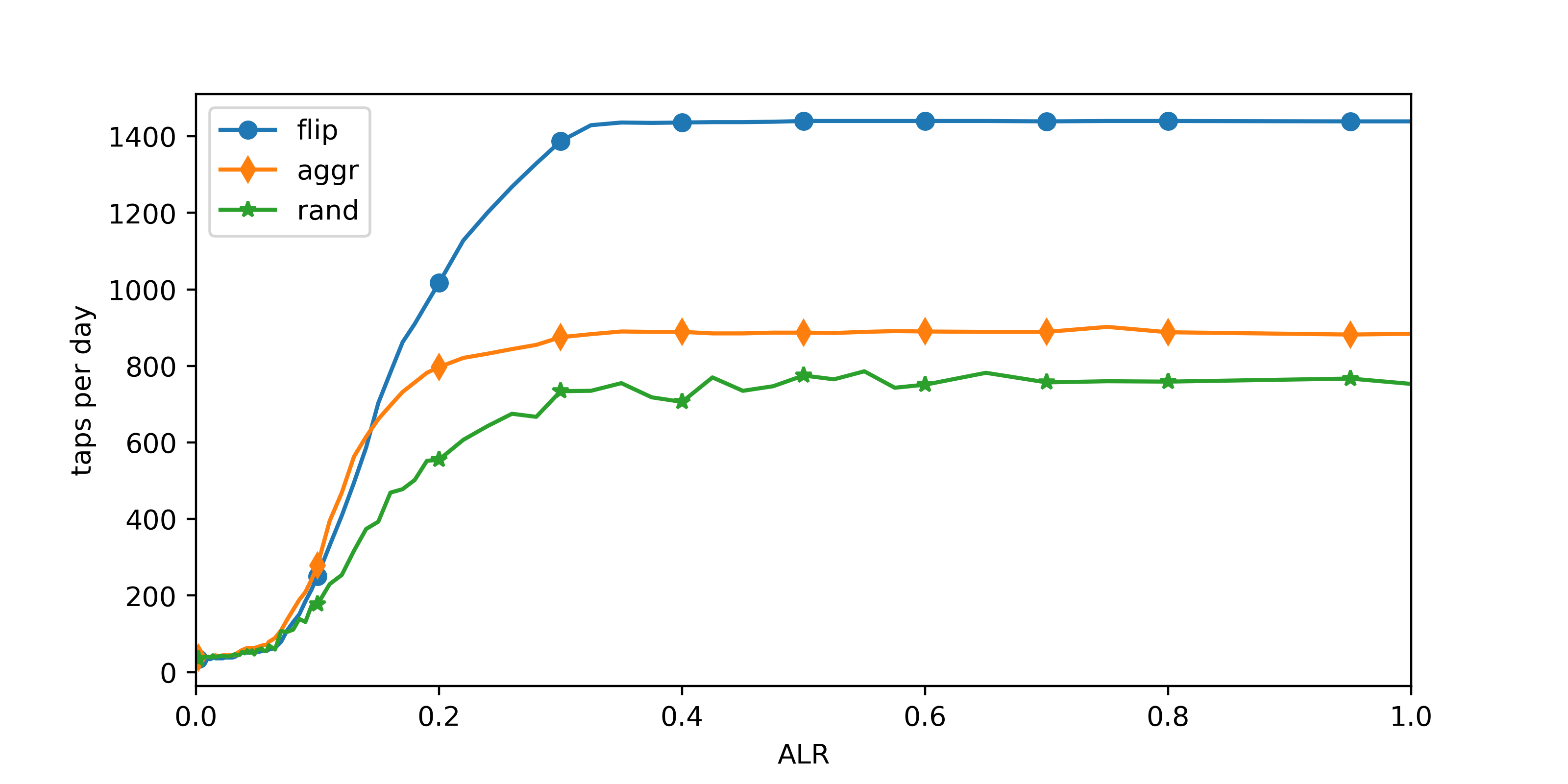}
  \vspace{-10pt}
  \caption{Number of OLTC tap changes within a simulated day as a function of attacker load ratio (ALR) and attack strategy, for the flip, heuristic, and random attacks, respectively.}
  \Description[Three lines showing how OLTC taps per day increeases with ALR.]{Three line graphs, with the X axis "ALR" ranging from 0.0 to 1.0, and the Y axis "taps" ranging from 0 to 1440. Each of the three lines increase from around 35 taps per day, following roughly S shaped curves, and reaching different asymptotes as ALR approaches infinity.. The "random" line increases from 35, following a bumpy S-shape, reaching approximately 720 taps per day by ALR 0.30. The "aggr" method increases to roughly 800 taps per day by ALR 0.30. The "flip" method increases to roughly 1440 taps per day by ALR 0.30.}
  \label{fig:401_taps_vs_ALR}
\end{figure} 

\begin{figure}[h]
  \centering
  \includegraphics[width=.8\textwidth]{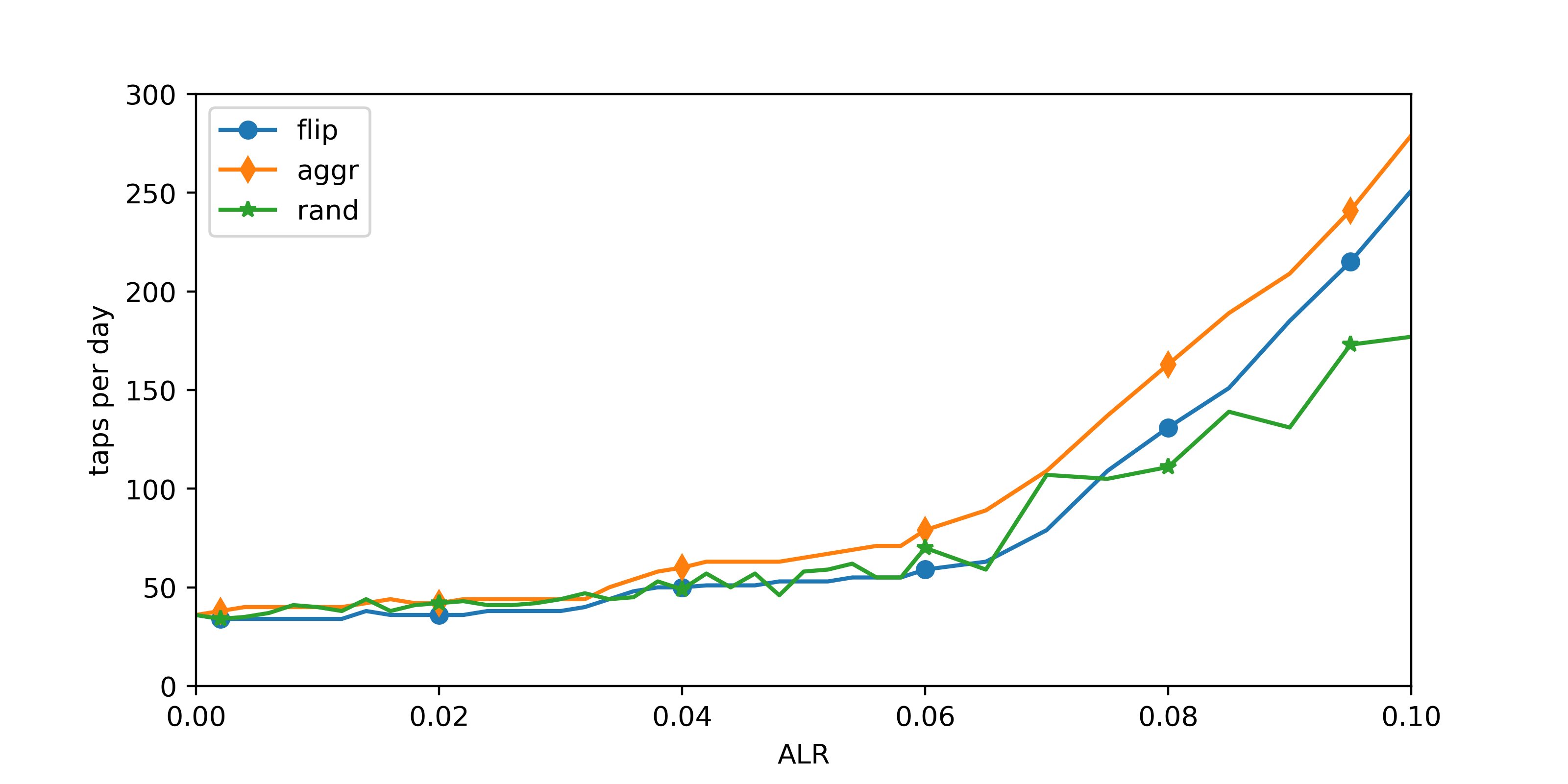}
  \caption{Number of OLTC tap changes within a simulated day as a function of attacker load ratio (ALR) and attack strategy, for lower values of ALR, for the flip, heuristic, and random attacks, respectively.}
  \Description[Three lines showing how OLTC taps per day increases with ALR.]{Three line graphs, with the X axis "ALR" ranging from 0.0 to 0.10, and the Y axis "taps" ranging from 0 to 400. Each of the three lines increase from around 35 taps per day, curving upwards. The "rand" method is a bumpy line increasing to about 150 taps per day for ALR 0.10. The "flip" method reaches about 200, while the "aggr" method is slightly higher, reaching about 300.}
  \label{fig:401_taps_vs_ALR_10p}
\end{figure} 

\begin{figure}[h]
  \centering
  \includegraphics[width=.8\textwidth]{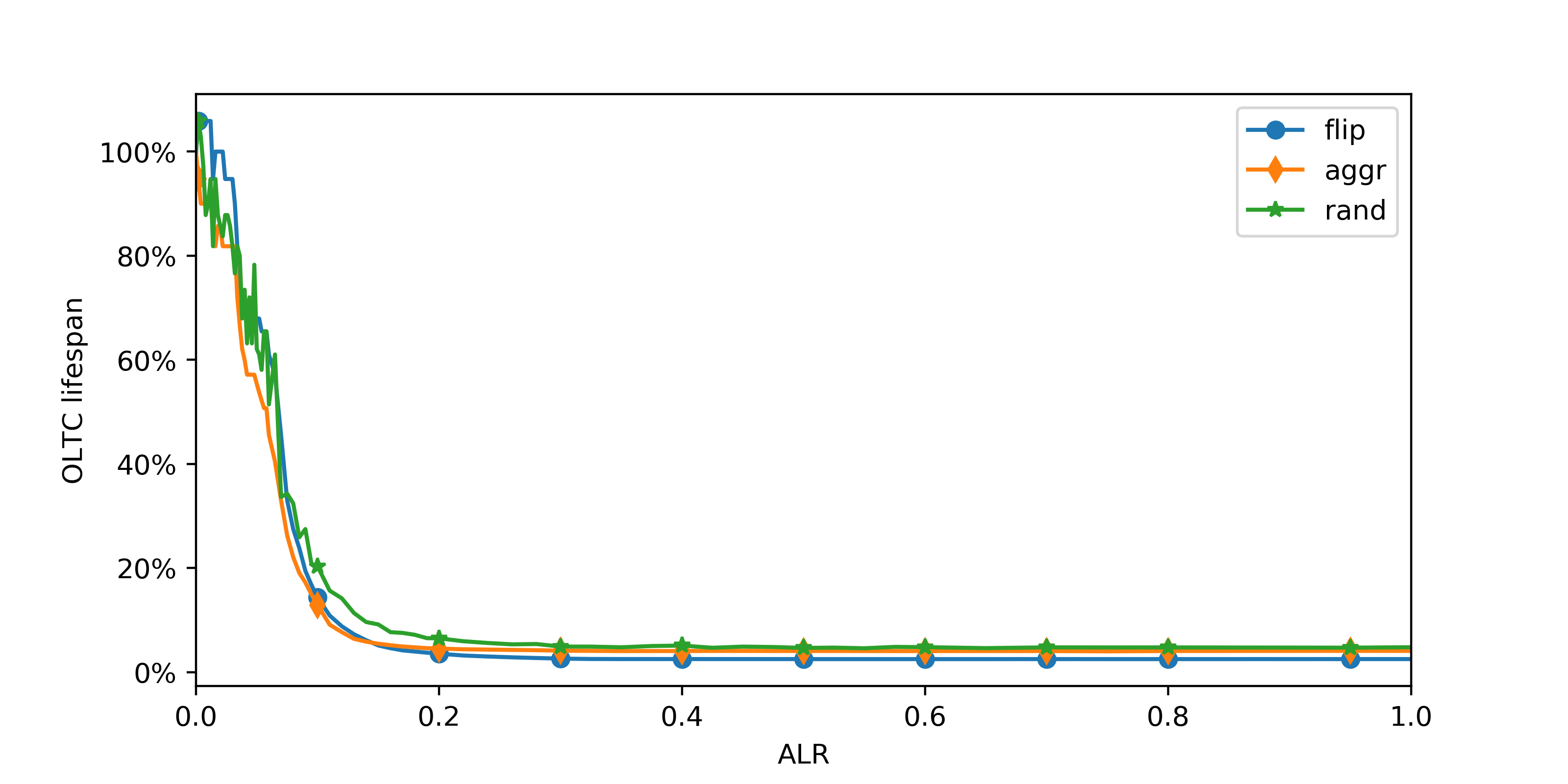}
  \vspace{-10pt}
  \caption{OLTC lifespan degrading as a function of attacker load ratio (ALR) and attack strategy, for the flip, heuristic, and random attacks, respectively.}
  \Description[Three lines showing a dramatic drop in OLTC lifespan as a function of attacker load ratio.]{Three line graphs, with the X axis "ALR" ranging from 0.0 to 1.0, and the Y axis "lifespan" ranging from 1.0 to 0.0. Each of the three lines drop from 1.0 to less than 0.10 for ALR values between 0.0 to 0.20.}
  \label{fig:401_lifespan_vs_ALR}
\end{figure}

\begin{figure}[h]
  \centering
  \includegraphics[width=.8\textwidth]{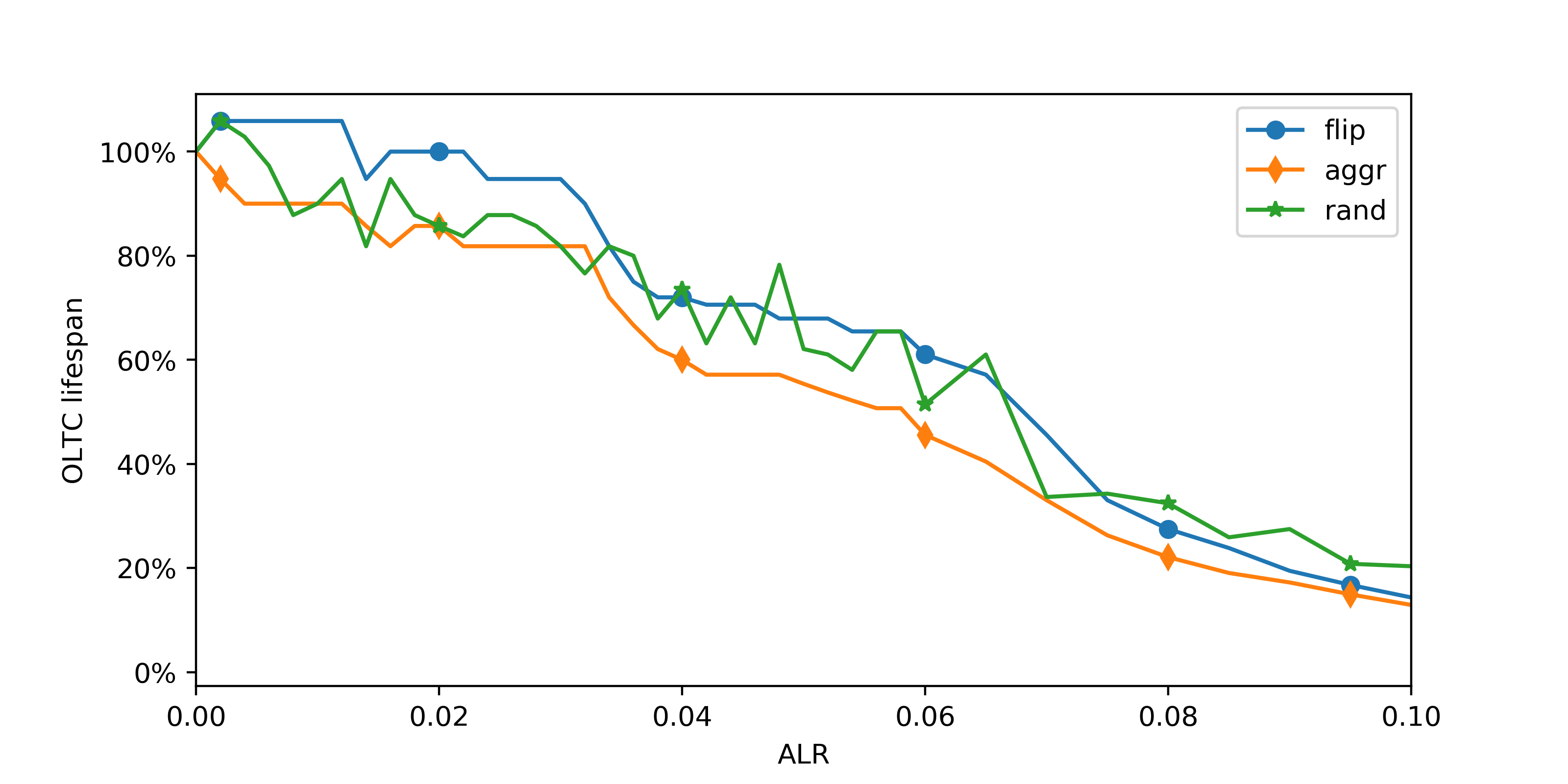}
  \caption{OLTC lifespan degrading as a function of attacker load ratio (ALR) and attack strategy, focused on the lowest values of ALR, for the flip, heuristic, and random attacks, respectively.}
  \Description[Three lines showing how OLTC lifespan decreases with ALR.]{Three line graphs, with the X axis "ALR" ranging from 0.0 to 0.10, and the Y axis "lifespan" ranging from 1.0 to 0.0. Each of the three lines drop from 1.0 to around 0.20 for ALR values between 0.0 to 0.10.}
  \label{fig:401_lifespan_vs_ALR_10p}
\end{figure} 

In Figure~\ref{fig:401_taps_vs_ALR} we see how the three different attack strategies perform in terms of causing tap changes. The ``flipping" strategy is the most effective, causing $1440$ tap changes (the maximum possible in a given simulation) for $ALR \geq 0.50$. In Figure~\ref{fig:401_lifespan_vs_ALR} we see the inverse, and the figure shows how quickly the OLTC lifespan drops with increased $ALR$, showing a reduction to less than $10\%$ of original lifespan for an $ALR \geq 12\%$. We focus on values of $0 \leq ALR \leq 0.10$ in Figure~\ref{fig:401_taps_vs_ALR_10p} and in Figure~\ref{fig:401_lifespan_vs_ALR_10p}, showing how the heuristic strategy performs slightly better than the flipping strategy for low values of $ALR$. Surprisingly, for very low values of $ALR$, the flipping strategy actually results in a slightly \textit{increased} OLTC lifespan, by causing less tap changes than is usual.

\textcolor{black}{These reduction in lifespan figures exemplify the increased costs of maintaining the expensive OLTC transformer. The increased rate of taps result in a proportionally increased maintenance cost and an inversely proportional decrease in lifespan.}

\textcolor{black}{From the mechanics of the OLTC, it is not surprising that the flipping strategy generally outperforms the other three strategies. The OLTC has a switch that moves in response to changes in demand, and the flipping strategy causes such changes at the fastest frequency and amplitude possible in our simulator. The heuristic attack is meant to flip only when it is expected to cause a tap change, but the heuristic is not perfect. One reason that the heuristic attack only beats the flipping attack at low values of ALR might be that the flipping attack incidentally \textit{prevents} a tap change by canceling out demand dynamics, whereas the heuristic attack only activates when it is expected to cause a change. This is best exemplified in Fig.~\ref{fig:401_lifespan_vs_ALR_10p}, which shows the flipping attack actually increasing lifespan for very low values of ALR.}



\section{Learning Based Localization Problem Formulation} \label{sec:learning_problem}

In this section, we describe the data generation process for learning and the localization goal and the learning problem. Without a valid model that can be used for identifying the locations of power bots, in this and the following section we approach localization entirely as learning problem. Without the presence of a simulated attacker, a simulation as generated in OpenDSS has no stochastic elements and so is deterministic. With fixed household demand data as generated by CREST, the attacker is the only factor producing meaningfully different scenario samples. So, we formulate our learning problem as the supervised task to identify which set of nodes contain an attacker conditioned on an attacker already being present in the system. Specifically, this is formulated as a supervised classification problem with thirteen classes, one for each node in the neighborhood considered. (This node-level localization is limited by the fidelity of the simulation and household demand data. A more precise simulation or real-world scenario could perform localization at the neighborhood level or even the household level.)  

\textcolor{black}{With this formulation, in a real-world setting, grid operators can actually gather labelled training data by simulating these attacks against a power system model. Many grid operators already have a `digital twin' simulation of their system so ~\cite{tao2018digital, jain2019digital, zhou2019digital}. For instance, there has been methods using digital twin to estimate the measurable characteristic outputs of a PV energy conversion unit in real time to perform fault diagnosis~\cite{zhou2019digital}.
Grid  operators  often  deploy  a  ‘digital  twin’  model  of  their  real  system. Assuming that distribution  grid  operators  have  the  capacity  to  readily  use  the  model of the system based on digital twin or similar simulators, labeled training data under attacks can be simulated and generated, and then used for learning-based attack localization methods.}

In Table~\ref{tab:problem-notation} we define the notation used in this and the following section.

\label{tab:problem-notation}
\begin{table}[h]
\caption{Learning Based Localization Problem Notations and Symbols}
\begin{center}
\begin{tabular}{r p{8cm} }
\toprule
    $n \in 1..N$    & Simulation index; \\ 
    $t \in 1..T$    & Discrete 1-minute time index;\\ 
    $k \in 1..K$    & Distribution grid node index; \\ 
    $d$             & Dimensionality of grid state snapshot;\\ 
    $X$             & Simulation dataset, with shape  $(N, T, d)$\\
    $Y$             & Simulation attacker data, with shape $(N, T, k)$.\\
    $w$             & Time-window horizon, must cleanly divide $T$.\\
    $X^w$           & Reshaped sample dataset, with shape $(NT/w, dw)$\\
    $Y^w$           & Reshaped target dataset, with shape $(NT/w, K)$\\
\bottomrule
\end{tabular}
\end{center}
\end{table}

We use the example when there are $K=13$ nodes in focus can be attacked to explain the learning problem. This means there are $2^{13} = 8192$ possible combinations of nodes. Due to the large dimensionality of a single simulation scenario, we only simulate a subset of scenarios, considering scenarios where an attacker has $1$, $2$, $11$, or $12$ nodes, which results in ${13 \choose 1} + {13 \choose 2} + {13 \choose 11} + {13 \choose 12} = 182$ different combinations. Compared to all $2^{13}=8192$ combinations, this choice of values reduces the size of data while maintaining a balanced dataset. \textcolor{black}{We do not expect it to be necessary to generate samples for all $2^k$ scenarios in order to get sufficient learning results. The computational costs should scale roughly linearly with the number of output nodes, and we don't expect this learning formulation to scale beyond the reach of modern computational resources. This is because we formulate our problem as a multi-label binary classification problem. So, while there are $2^k$ output combinations, there are only $k$ output neurons, i.e. $k$ bits or $R^k$ output dimensionality. Hence, while it is true that there are $2^k$ potential binary outputs, we don't expect any of the models we use to scale with $2^k$ in terms of training time, inference time, or storage. The computational costs should scale roughly linearly with output node size.}

We also consider the three attack strategies and four values for ALR, $0.050, 0.075, 0.100,$ and $0.125$, resulting in $N = 182 * 3 * 4 = 2184$ simulated scenario samples. Each simulation represents one day, with $T = 1440$ minute-long time steps. Power grid state data captured at each time step is a vector in $\mathbb{R}^d$, where dimensionality $d = 791$. This vector contains the tap values of OLTCs, as well as the vector components of voltage, current, reactive power and inductive power of OLTCs and capacitors in the power grid. We call this dataset $X$ with shape $(N, T, d)$.

For simulation, we also have 1440 samples with labels of dimensionality $K=13$, which have a binary value $0$ or $1$, with $1$ active if an attack has a device active (``on" and consuming power) in the at node. This label dataset $Y$ is of shape $(N, T, K)$.

We transform the sample dataset $X$ using non-overlapping time windows of size $w$, where $w$ is a positive integer that divides $T$. We call these transformed dataset $X^w$ with shape $(N * T / w, d * w)$. So, there are $NT/w$ learning samples in $X^w$ of dimensionality $dw$. We transform the target dataset $Y$ as well, into a dataset labeled $w$. Each sample is of dimensionality $K$, each value $1$ if the node was active at any period in the window $w$ and $0$ if not. So, it has shape $(N * T / w, K)$, meaning there are $NT/w$ target values of dimensionality $K$.

For the training dataset, we permute the samples before transformation, and use a 50:50 train-test split for the transformed data with cross-validation. The learning objective is to maximize the accuracy when mapping samples in $X^w$ to $y^w$. Specifically, a trained classifier produces a prediction $\text{clf}(x_{test}) = \hat{y} \in \{0,1\}^{13}$ where $\text{clf}$ is a trained classifier, $x_{test}$ is some testing sample, and $\hat{y}$ is some prediction. The goal is to minimize the number of mispredictions, that is, to minimize $$\sum_{x_{test}, y_{test} \in \left(X^w_{test}, Y^w_{test}\right)} \sum_{i = 1}^{13} \left|\hat{y}_i - y_{test, i}\right|,$$ where $(X^w_{test}, Y^w_{test})$ is the set of pairs of testing samples and targets.

\section{Learning-Based Localization Method and Results} \label{sec:learning_metroid}

\textcolor{black}{In this section, we employ experiments on three different methods--- SVM, neural network, and decision tree, according to the parameters defined in the previous section. We evaluate these methods performance in properly identifying which of the 13 nodes contain attacking devices. All of these models train to functions with inputs of dimensionality $d\cdot w$ and outputs of dimensionality $k$, so they are expected to scale reasonably. We use 50:50 train-test split for the simulated data with cross-validation and show the evaluation results in figures.}


\subsection{Evaluation Metrics}

\newcommand\TP{\mathit{TP}}
\newcommand\TN{\mathit{TN}}
\newcommand\FP{\mathit{FP}}
\newcommand\FN{\mathit{FN}}

We evaluate the localization method as thirteen separate binary classification problems. We utilize three scores for evaluation: Normalized accuracy, true positive rate (sensitivity), and true negative rate (specificity). The performance of a classifier against the test set is described by the number of true negatives $\TN$, false positives $\FP$, false negatives $\FN$, and true positives $\TP$. Let $N_P = \FN + \TP$ be the total number of positive samples in the test set and $N_N = \FP + \TN$ be the total number of negative samples in the test set.

The ``normalized accuracy" is an improved measure of performance versus accuracy in the case of imbalanced data. \textcolor{black}{It compares the measured accuracy against a ``baseline accuracy" which would be achieved by a model which guesses the majority class from a training set, in essence measuring our accuracy improvement from this simple threshhold-based classifier.} The \emph{accuracy} is calculated as $\mathit{acc} = \frac{\TP+\TN}{N_P + N_N}$. The \emph{baseline accuracy} is calculated as $\mathit{acc_B}=\frac{\max{\left(N_P,N_N\right)}}{\TN+\FP+\FN+\TP}$. From these two values, we calculate the \emph{normalized accuracy} as $$\mathit{normacc} = \frac{\mathit{acc}-\mathit{acc_B}}{1-\mathit{acc_B}}.$$ 
A value of $normacc = 0$ represents no improvement from the baseline, whereas $normacc = 1$ represents perfect accuracy. This calculation is similar to but distinct from Cohen's Kappa~\cite{mchugh2012interrater}.

\subsection{Decision Tree Classifier Results and Analysis}

\begin{figure}[h]
  \centering
  \includegraphics[width=.8\textwidth]{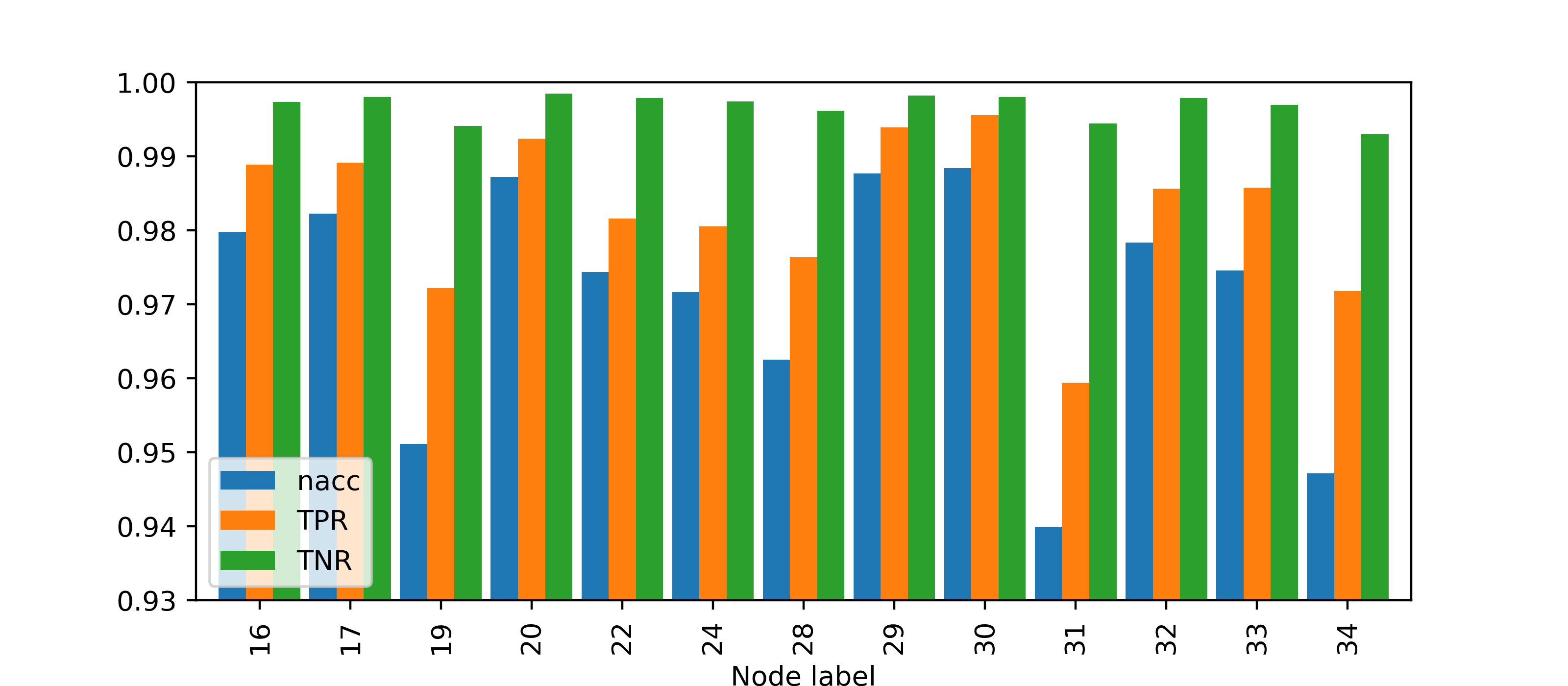}
  \caption{Localization metrics (normalized accuracy,  true positive rate / sensitivity, and true negative rate / specificity) of the trained decision tree on the test set for each of the 13 nodes. Note the y-axis starts at $0.93$.}
  \Description[The decision tree performs very well on the test set.]{Thirteen sets of three bar graphs, labelled for each of the thirteen targeted nodes. Among the thirteen graphs, the normalized accuracy varies between about 94\% and 99\%, the true positive rate varies between about 96\% and 99\%, and the true negative rate varies between about 98\% and 100\%.}
  \label{fig:351_detection-metrics}
\end{figure}

Because no well-defined model exists that could allow us to analytically perform localization, we address the localization problem by a learning-based approach. We detail our experiments here. We apply the decision tree classifier with default parameters as provided by Scikit Learn 0.21.3~\cite{scikit-learn}, utilizing a time window of $w=1$ to the $50\%$ training split of the dataset. Because $w=1$, this means the classifier works on instantaneous ``snapshots" of the grid state and does not consider time-series dependencies.\footnote{The default DecisionTreeClassifier uses the Gini impurity as the splitting criterion~\cite{gelfand1989iterative} and splits according to the highest-value criterion (denoted ``best" strategy by SciKit Learn.) There is no set maximum depth nor set maximum number of leaf nodes, the minimum samples per leaf is None, and the minimum samples per leaf is one.} The results of the decision tree classifier as applied to the testing set are provided in Figure~\ref{fig:351_detection-metrics}. Specifically, a trained classifier takes as input values the power-grid data in $\mathbb{R}^d$ and outputs the target values in $\{0, 1\}^{13}$.

We see high-performance results in Figure~\ref{fig:351_detection-metrics} without any tuning or time-series analysis, with normalized accuracies between $93\%$ and $99\%$, a true positive rate between $96\%$ and over $99\%$, and a true negative rate between $99\%$ and $100\%$.

Of most importance is the normalized accuracy, which directly reflects the model improvement from a baseline majority-rule classifier. Compared with more popular methods such as neural networks and SVMs (discussed below), this suggests that decision trees have strong potential for localization in the context of power botnet attacks.

The normalized accuracy also varies between the different nodes, with the model performing the worst on nodes 31, 34, and 19, and performing the best on nodes 20, 29, and 30. Because nodes 31, 34, and 19 are internal nodes while 20, 29, and 30 are edge nodes, this suggests internal nodes are most resistant to localization. (See to Figure~\ref{fig:IEEE123}.)

\subsection{Comparison to other learning-based methods}



In this section, we briefly discuss other learning methods and analyze their different performances. SVM-based and neural network based learning methods are both widely used in the areas of machine learning and AI, and were among the first models we tried. \textcolor{black}{For the SVM based methods, we used linear SVM classifiers with a maximum of 1000 training iterations.  Similar to the decision tree classifier, all other SVM parameters were the defaults provided for svm.LinearSVC in  Scikit Learn 0.21.3. For neural network based methods, we tried a variety of hyperparameters and architectures (including LSTM~\cite{hochreiter1997long} for the $w=60$ case and multi-layer perceptrons of various widths and depths.) The most efficient neural network architecture we found was a simple linear perceptron with sigmoid activation and window size $w=1$.} 

\begin{figure}[h]
  \centering
  \includegraphics[width=.9\textwidth]{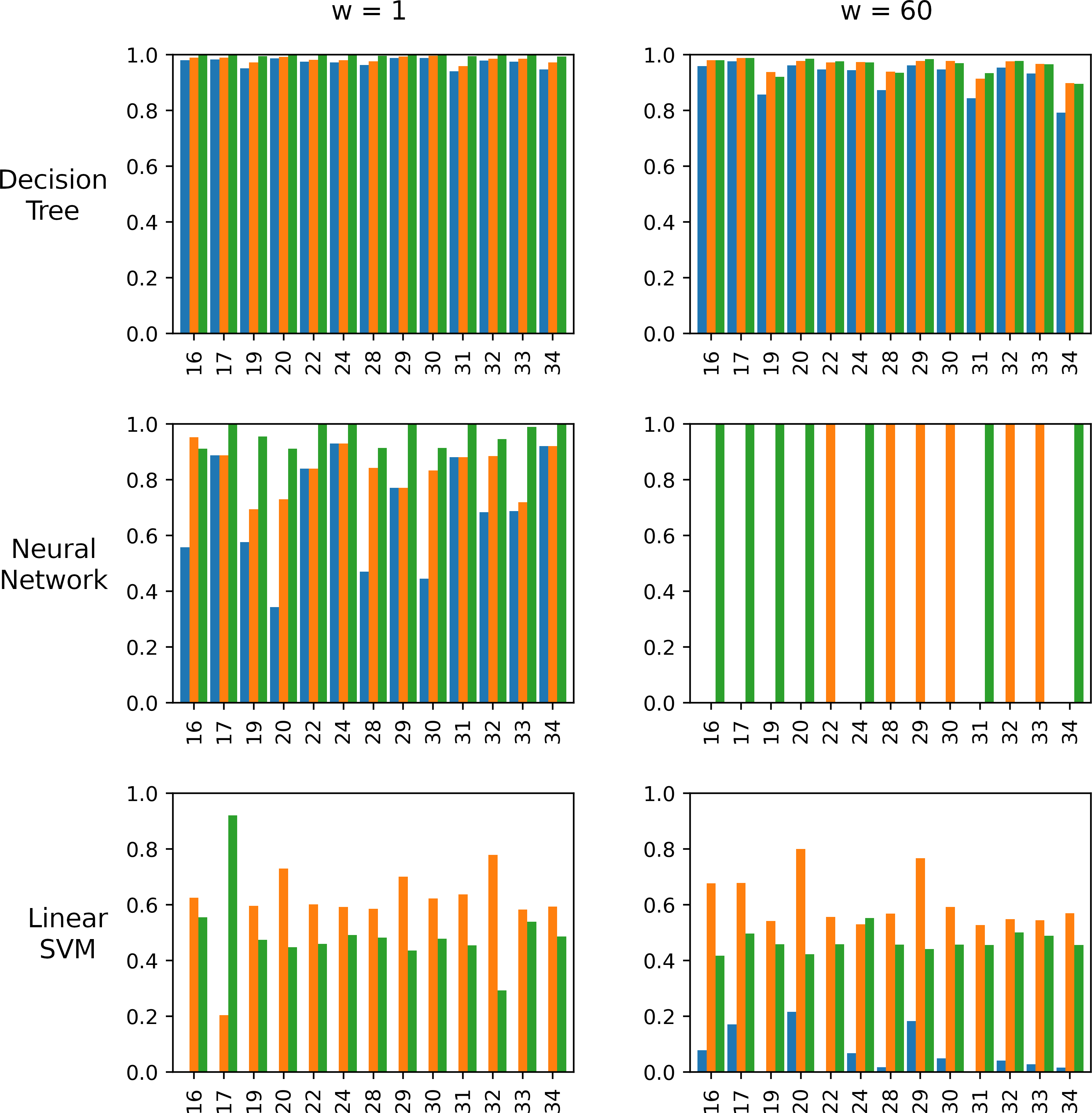}
  \caption{Comparison of the impact of learning method and of window size $w$ on classifier performance. Vertically, we see the performance of a decision tree classifier, a best-effort neural network classifier, and a linear SVM classifier. Horizontally, we see the impact of different choice for window size $w$. Here we see decision trees perform the best, especially so when $w = 1$.}
  \Description[Decision trees with $w=1$ perform the best.]{Six graphs, of the same exact  format as that in \ref{fig:351_detection-metrics}, arranged in a three-by-two grid. From top to bottom, we compare three methods (decision tree, neural network, and linear SVM. The leftmost column are results for $w=1$, and the rightmost column are results for $w=60$. Let's now compare the graphs by inspecting which have the highest (i.e. best performing) normalized-accuracy columns. We see Decision Tree with $w=1$ (upper-left graph) perform the best, with Decision Tree with $w=60$ (top-right) a close second with slightly lower performance. Neural networks with $w=1$ come in third (middle-left), and SVM with $w=60$ comes in fourth (bottom right). } 
  
  \label{fig:440-negative-results}
\end{figure} 

\paragraph{Impact of $w > 1$} When working with time series data, the window size chosen for data samples (denoted here as $w$) is important. When training models on window sizes $w > 1$, we were surprised to see performance drop in the case of decision trees and neural networks. This meant that we saw the best performance with $w=1$, meaning our models took no advantage of time series characteristics. This difference in performance is explained by the ``curse of dimensionality", a phenomena where machine learning models fail with higher dimensional data~\cite{trunk1979problem}. Choosing $w > 1$ results in higher-dimensional samples, and also lowers the number of training samples proportionally. For the decision tree $w=60$ case, the normalized accuracies range between 79\% and 97\%.

The results are surmised in Figure~\ref{fig:440-negative-results}. There, we compare the performance of our decision trees, our best-attempt neural networks, and our linear SVM  based classifiers, and their performance for values $w = 1$ (i.e. instantaneous data) and $w = 60$ (i.e. samples are entire hours worth of data.) The normalized accuracy of the neural network $w=1$ case range between 34\% and 93\%, and the normalized accuracies of the SVM $w=60$ case range between less than -4\% and 22\%.

\section{Conclusion} \label{sec:conclusion}

This work is motivated by the need to secure one important cyber-physical systems (CPS) for our society--the power grid. Traditional cyber-attacks against power grids usually focus on compromising the internal SCADA control systems or denying service and access. In contrast, in this work we consider the capacity for an attacker to compromise consumer devices which exist at the edge of the power grid and use them to perform load altering attacks, without any need to target the internal SCADA control systems. We define the mechanism of \emph{power botnets} and how they are used to perform load altering attacks. We detail a simulated distribution-grid environment utilizing the OpenDSS simulator, the IEEE 123 test feeder topology, household-level power demand data provided using the CREST tool, and an analytic water heater model. We consider three strategies for an \emph{OLTC weardown attack} and show that even weak attackers drastically reduce the lifespan of OLTC transformers--with only $5\%$ load ratio, the OLTC lifespan is reduced to under $60\%$. The feasability of these attacks will increase as consumers continue to adopt high-wattage and inconsistently-secured IoT devices. Although we only model water heaters, other devices such as refrigerators, air conditioners, and desktop computers can be used in such an attack and we will analyze the effects as one of our future work. We then define  a supervised learning based problem to identify and locate such OLTC weardown attacks. We compare the performance of decision trees, neural networks, and SVMs, and we find decision trees to perform well with much less computational cost or parameter-tuning effort. Such a decision tree locates power bots with very high accuracy, with the classification true positive rate (sensitivity) exceeding $95\%$ and true negative rate (specificity) exceeding $99\%$. In the future, we will analyze the performance of unsupervised learning based attack localization methods that are well-suited to an intrusion detection environment, \textcolor{blue}{and utilize a reinforcement-learning based approach based to explore more attack strategies.}

\begin{acks}
This project was supported in part by the National Science Foundation under Grant ECCS-2018492. 
\end{acks}

\bibliographystyle{ACM-Reference-Format}
\bibliography{biblio}

\end{document}